\pdfoutput=1
\documentclass{article}
\usepackage{fancyvrb}
\usepackage[final]{neurips_2023}

\usepackage[utf8]{inputenc} % allow utf-8 input
\usepackage[T1]{fontenc}    % use 8-bit T1 fonts
\usepackage[hidelinks]{hyperref} 
\usepackage{url}            % simple URL typesetting
\usepackage{booktabs}       % professional-quality tables
\usepackage{amsfonts}       % blackboard math symbols
\usepackage{nicefrac}       % compact symbols for 1/2, etc.
\usepackage{microtype}      % microtypography
\usepackage{xcolor}         % colors
\usepackage[hang,flushmargin]{footmisc}
\usepackage{enumitem}
\usepackage{xspace}
\usepackage{arydshln}
\usepackage{multirow}
\usepackage{graphicx}

\usepackage{etoolbox}
\makeatletter
\patchcmd{\@verbatim}
  {\verbatim@font}
  {\verbatim@font\small}
  {}{}
\makeatother

\newcommand{\ignore}[1]{}
\newcommand{\mytt}[1]{\texttt{\small #1}}

\title{End-to-End Retrieval with Learned\\ Dense and Sparse Representations Using Lucene}

\author{Haonan Chen,$^{1}$ Carlos Lassance,$^{2}$ and Jimmy Lin$^{3}$ \\[1ex]
$^{1}$ Department of Electrical and Computer Engineering, University of Waterloo\\
$^{2}$ Naver Labs Europe \\
$^{3}$ David R. Cheriton School of Computer Science, University of Waterloo\\
}

\begin{document}
\maketitle

\begin{abstract}
The bi-encoder architecture provides a framework for understanding machine-learned retrieval models based on dense and sparse vector representations.
Although these representations capture parametric realizations of the same underlying conceptual framework, their respective implementations of top-$k$ similarity search require the coordination of different software components (e.g., inverted indexes, HNSW indexes, and toolkits for neural inference), often knitted together in complex architectures.
In this work, we ask the following question:\ What's the simplest design, in terms of requiring the fewest changes to existing infrastructure, that can support end-to-end retrieval with modern dense and sparse representations?
The answer appears to be that Lucene is sufficient, as we demonstrate in Anserini, a toolkit for reproducible information retrieval research.
That is, effective retrieval with modern single-vector neural models can be efficiently performed directly in Java on the CPU.
We examine the implications of this design for information retrieval researchers pushing the state of the art as well as for software engineers building production search systems.
\end{abstract}

\section{Introduction}

The popular bi-encoder architecture has emerged as the dominant conceptual framework around which retrieval models are organized today.
This design formulates retrieval in terms of comparisons between vectors derived from queries and pieces of content (passages, documents, and even images and multimedia).
More precisely, the core retrieval problem can be framed as nearest neighbour search:\ given a query vector, the system's task is to return the $k$ ``closest'' content vectors, typically in terms of a simple operation such as the inner (dot) product.

Central to the bi-encoder design is representation learning, where deep learning techniques are typically applied to learn vector representations that maximize the inner product between queries and relevant content while minimizing the inner product between queries and non-relevant content.
At a high-level, these vector representations can be categorized as {\it dense}, typically contextual representations derived from transformers~\citep{Vaswani_etal_NIPS2017} that capture latent semantics, or {\it sparse}, where the dimensions of the vectors are defined by the vocabulary space.
Thus, we arrive today at retrieval models based on either learned dense or learned sparse representations.

Within this conceptual framework, we focus on the software infrastructure necessary to realize end-to-end retrieval with learned dense and sparse representations.
In particular, we ask:\ What's the simplest design that could ``work''?
We operationalize this in terms of minimality, or requiring the smallest number of changes to existing, established infrastructure.
Our answer might be somewhat surprising, at least to some:\
We find that Lucene is sufficient.
Inverted indexes using the ``fake words'' trick enable retrieval with learned sparse representations.
Lucene's HNSW indexes enable retrieval with learned dense representations.
Finally, with the integration of the ONNX Runtime, query inference can be performed directly from Lucene.
This means that effective end-to-end retrieval with modern neural models can be efficiently performed in Java, on the CPU.
In particular, we were able to ``pare away'' two components that many have perhaps thought to be indispensable to neural retrieval today:\ a separate vector store to enable top-$k$ retrieval on dense vectors and Python as a bridge to neural inference engines.

We demonstrate end-to-end retrieval with learned dense and sparse representations in Anserini~\citep{anserini}, an existing Lucene toolkit for reproducible information retrieval research.
This design holds implications for information retrieval researchers as well as for engineers who build production search systems.
For researchers, simplicity accelerates progress by lowering hurdles and streamlining research iteration cycles.
For system builders, Anserini primarily provides a prototype of what's possible:\
Since Lucene resides at its core, anything that can be implemented in Anserini can just as easily be integrated into Elasticsearch, OpenSearch, Solr, or any other software component in the Lucene ecosystem.
This suggests that enterprises can take advantage of the latest advances in neural retrieval models without abandoning existing infrastructure investments (e.g., the ELK stack).
Capabilities that are already present in Lucene provide a compelling adoption path.

\section{Background and Related Work}

\begin{figure*}[t]
\begin{center}
\centerline{\includegraphics[width=0.6\textwidth]{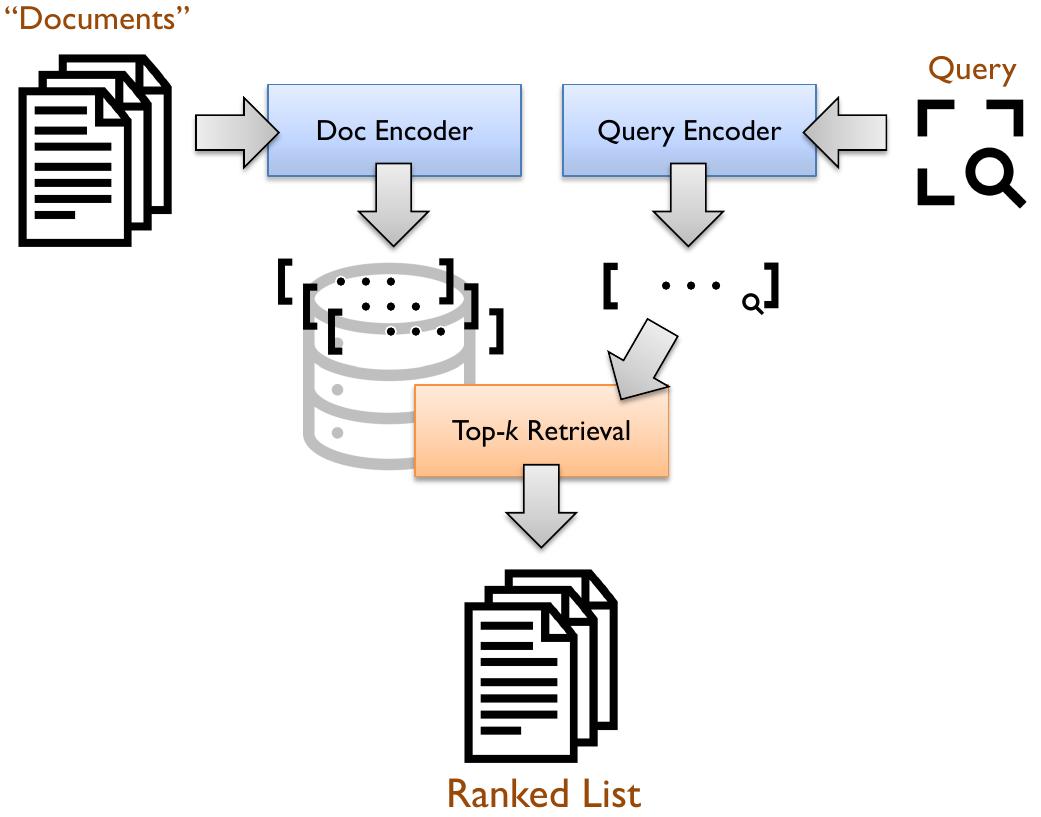}}
\caption{Illustration of the bi-encoder architecture, where encoders generate vector representations from queries and documents. In this design, retrieval is formulated as $k$-nearest neighbour search in vector space.} 
\label{bi-encoder}
\end{center}
\end{figure*}

An illustration of the bi-encoder architecture is shown in Figure~\ref{bi-encoder}, where retrieval is framed as nearest neighbour search in vector space.
This design assumes the existence of encoders that convert documents and queries into vector representations.
Note that throughout this work we use ``documents'' in a generic sense to refer to content that we wish to search, even though in actuality they may represent passages, web pages, or even images and other multimedia.

Given the vector representation of a query, the system's task is to efficiently retrieve the top-$k$ most similar document vectors with respect to a simple operation such as the inner (dot) product.
\citet{lin2022proposed} observed that in this framework, there are two important characteristics of these representations:

\begin{itemize}[leftmargin=0.5cm]

\item the basis of the vector space---for example, dense vectors where the dimensions capture latent semantics, or sparse vectors where the dimensions are defined by the vocabulary;

\item whether the vectors are generated by a machine-learned model (typically in a supervised manner) or via some heuristic weighting function (typically in an unsupervised manner).

\end{itemize}

\noindent Cast in this framework, ``traditional'' bag-of-words BM25 retrieval~\citep{Robertson_Zaragoza_FnTIR2009} can be characterized as sparse (lexical) representations generated by a heuristic weighting function (that is, the BM25 scoring function).
On the other hand, dense retrieval models such as DPR~\citep{dpr} can be characterized as learned dense representations.
These models, along with vector stores for manipulating dense vectors, dominate today's discourse.

\subsection{The Vector Store Narrative}

Let us begin by recounting the ``standard narrative'' behind ``semantic search'' today.
The story, parroted endlessly on social media, blog posts, and in other venues, typically begins with the awesome power of transformers to generate dense (semantic) representations that are able to capture the ``meaning'' of text.
Such learned dense representations, or embeddings, are supposedly light-years ahead of bag-of-words lexical representations in terms of retrieval effectiveness.
The truth is far less hyperbolic, as dense retrieval models still underperform traditional lexical models in at least some scenarios~\citep{beir,sciavolino-etal-2021-simple,Kamalloo_etal_arXiv2023_BEIR}.
Embeddings are also touted as the key ingredient to retrieval augmentation, a popular method of enhancing the capabilities of large language models (LLMs)~\citep{Mialon:2302.07842:2023,asai-etal-2023-retrieval}.

But there's one catch:\ top-$k$ retrieval with dense vector representations requires a completely new software stack built around ``vector stores''.
Listening to the breathless hype surrounding these new-fangled contraptions,  one could easily imagine the obsolescence of the venerable inverted index, the workhorse on which bag-of-words retrieval models rely.
One goal of this paper, to butcher a quote often (mis-)attributed to Mark Twain, is to push back against this narrative:\ ``reports of the death of inverted indexes are greatly exaggerated''.

While learned dense representations have dominated social media and blog posts---with countless startups peddling vector stores---many researchers have been developing learned {\it sparse} representations in (relative) obscurity, e.g.,~\cite{Dai_Callan_SIGIR2020,Bai:2010.00768:2020,gao-etal-2021-coil,Mallia_etal_SIGIR2021,unicoil,nguyen2023unified}, just to cite a few.
This class of models takes advantage of the same innovations that have fed learned dense retrieval models, but instead generate sparse vectors that are grounded in the vocabulary space.
With recent advances, models such as SPLADE++~\citep{spladepp} have achieved retrieval effectiveness at least on par with dense retrieval on a variety of standard benchmarks~\citep{Kamalloo_etal_arXiv2023_BEIR}.

Retrieval with sparse representations can be implemented with inverted indexes using the widely-known ``fake words'' trick~\citep{Mackenzie_etal_TOIS2023}, which means that they are compatible with existing libraries such as Lucene.
This has already been demonstrated in the Anserini IR toolkit~\citep{anserini} built around Lucene, which has incorporated sparse retrieval models since 2021~\citep{unicoil}.
Furthermore, \citet{devins2022aligning} showed that what's possible in Anserini is equally possible in Elasticsearch and any other component in the broader Lucene ecosystem.

In truth, {\it both} dense and sparse representations are indispensable, as the combination of evidence from both approaches leads to higher-quality output than either alone~\citep{Ma_etal_arXiv2021_DPR,Kamalloo_etal_arXiv2023_BEIR}.
Yet, the fact remains that top-$k$ retrieval with dense and sparse representations require different machinery:\ while sparse representations are amenable to inverted indexes, current best practices recommend the use of hierarchical navigable small-world network (HNSW) indexes for top-$k$ retrieval on dense representations~\citep{HNSW}.
However, \citet{Lin_etal_arXiv2023_OpenAI} argued that the need for the capabilities offered by HNSW indexes doesn't necessarily require full-blown vector stores as independent software components.
In fact, the authors pointed out that Lucene provides implementations of HNSW indexes\footnote{\scriptsize{\url{https://cwiki.apache.org/confluence/display/LUCENE/Release+Notes+9.0}}} that are adequate for servicing dense retrieval models.
They argued that, from a simple cost--benefit analysis, there does not appear to be a compelling need for dedicated vector stores, since vector search capabilities already exist in Lucene.
In this sense, both techniques are converging, with Lucene adding vector search and most vector stores developers starting to realize the need for basic search techniques, such as filters and inverted indexes.\footnote{\scriptsize{\url{https://news.ycombinator.com/item?id=37767033}}}

\subsection{The ONNX Runtime}

There is one final piece to the puzzle:\ retrieval with learned dense and sparse representations requires encoders for generating the representation vectors, typically using transformers.
This breaks down into two distinct processing phases:\ encoding the corpus and encoding the queries.
Encoding a (static) corpus is a one-time operation that demands high throughput and is usually impractical without GPUs.
We treat this as a form of preprocessing and assume that the entire corpus has already been converted into vector representations prior to ingestion into Lucene.
Query encoding, on the other hand, must be performed at search time, and unlike corpus encoding, is latency-sensitive.
Furthermore, it would be desirable if query encoding could be performed on the CPU, since such an implementation would result in significantly lower costs, better scaling characteristics (since it is easier to add CPU cores than GPUs), and expand the range of hardware that search applications can be deployed to.
This is exactly the scenario we explore:\ query encoding on the CPU.

When discussing neural inference, Python immediately comes to mind, as it has emerged as the dominant front-end language to popular neural toolkits such as PyTorch~\citep{pytorch}.
But Lucene is implemented in Java, which presents a language mismatch.
Previously, researchers have addressed this issue in different ways.
Pyserini~\citep{pyserini} provides Anserini bindings in Python, and thus for sparse retrieval models, query encoding can be performed using PyTorch before the query is passed to Anserini (via a Python-to-Java bridge) for top-$k$ retrieval.
Previously, without the benefit of PyTorch bindings via Python, Anserini by itself relied on pre-encoded queries for both dense and sparse retrieval with neural models.
That is, inference is applied to the queries ``outside the system'' and cached.
This reflects the scenario where a Lucene-based search engine calls, for example, an external inference API to perform query encoding.
While this is certainly an acceptable design pattern, it adds latency (from inter-process communication) and lacks the self-contained elegance that would come with an in-process query encoding solution (i.e., Java-based neural inference).

This is where ONNX (Open Neural Network Exchange), an open standard for computational graphs, comes in.
The ONNX Runtime provides cross-platform, cross-language, and cross-device support for neural inference---including using Java, the language of Lucene's implementation.
This means that with ONNX integration in Anserini, it is possible to perform retrieval with learned dense and sparse models {\it directly} in Java, on the CPU---that is, without needing PyTorch, a vector store, or even a separate HNSW library such as Faiss~\citep{faiss}.

\subsection{Putting Everything Together}

In other words, we find that Lucene is all you need for end-to-end retrieval with learned dense and sparse representations.
To be more precise, this work focuses on single-vector representations, setting aside multi-vector approaches such as ColBERT for dense representations~\citep{colbert} and SLIM for sparse representations~\citep{Li_etal_SIGIR2023}.
Lucene is the answer to the motivating question we posed:\ What's the simplest design, in terms of requiring the fewest changes to existing established infrastructure, that suffices to provide all our desired capabilities?

Although our experiments are conducted with Anserini, the capabilities we describe are easily exposed in Elasticsearch, OpenSearch, Solr, or any other component in the Lucene ecosystem.
Pyserini still provides Anserini bindings in Python, thereby offering seamless connections to other components of the modern Python-centric deep learning ecosystem.

To be clear, this is not the first work to examine or integrate ONNX into Lucene or any other search application.
In fact, our group has been experimenting with ONNX since 2017~\citep{tu-etal-2018-pay}.
More recently, ONNX has been integrated into Elasticsearch, currently as an experimental feature.\footnote{\scriptsize{\url{https://www.elastic.co/guide/en/machine-learning/master/ml-nlp-deploy-models.html}}}
Similarly, Vespa has integrated ONNX for some time now.\footnote{\scriptsize{\url{https://blog.vespa.ai/stateful-model-serving-how-we-accelerate-inference-using-onnx-runtime/}}}
However, the contribution of this paper is to ``put all the pieces together''.
To our knowledge, our implementation is the first that (1) is fully open-source, (2) covers both learned dense and sparse models, (3) provides end-to-end retrieval, including query encoding, (4) integrates with infrastructure that already has a massive install base---for the benefit of practitioners who focus on building production systems, and (5) provides support for formal evaluations on common IR benchmarks today---for the benefit of researchers who focus on model development.

\section{Design and Experimental Setup}

The core technical contribution of this paper is the integration of the ONNX Runtime into the Anserini IR toolkit to provide Java-based in-process query encoding.
To accomplish this, we implemented an abstract class called \mytt{OnnxEncoder}, which defines the (abstract) \mytt{encode} method for converting a string (the query) into a dense or sparse vector representation.
This abstract class has two sub-classes (also abstract), called \mytt{DenseEncoder} and \mytt{SparseEncoder}, for generating dense and sparse representations, respectively.
Concrete retrieval models instantiate one or the other class, as appropriate.
We experimented with an illustrative model from each class:

\begin{itemize}[leftmargin=0.5cm]

\item For the dense model, we selected the cosDPR-distil model~\citep{Ma_etal_CIKM2023}, primarily because Lucene's implementation of HNSW is restricted to top-$k$ retrieval using cosine similarity (as opposed to inner products in the general case).
The key difference lies in vector normalization, which most existing models do not perform.
Nevertheless, cosDPR-distil is trained according to best practices:\ an initial version of the model is used to mine hard negatives, which are then used to train a cross-encoder reranker and then distilled into the final retrieval model.

\item For the sparse model, we used SPLADE++ Cocondenser Ensemble-Distil~\citep{spladepp} (SPLADE++ ED for short).
As this model is well known, we simply refer interested readers to the cited reference for details.

\end{itemize}

To prepare the models for the ONNX Runtime, we first began by converting the PyTorch implementation of the encoders from HuggingFace~\citep{huggingface}.
We loaded the models using the HuggingFace API and then exported model weights using the \mytt{torch.onnx.export()} function available in PyTorch.
To further optimize the ONNX compute graphs extracted from the models, we used the \mytt{onnxruntime.transformers.optimizer} tool from the ONNX Runtime library, specifically developed for transformer models.
Note that further optimization is possible by quantizing weights to lower precision, but we decided to stay aligned with existing model weights and results.

As the focus of this work is query encoding at search time, we assumed that corpus encoding is handled separately and that we are provided with the raw representation vectors for ingestion into Anserini.
Our indexing implementation is based on the work described in~\citet{Ma_etal_CIKM2023}, except based on Lucene 9.8.0.
More details are provided in the results section.

We performed evaluations on the MS MARCO passage ranking test collection~\citep{msmarco}, a standard benchmark dataset used by researchers today.
Results are reported on the 6980 queries from the development (dev) set, as well as on queries from the TREC 2019 and 2020 Deep Learning Tracks~\citep{trec2019,trec2020}.
All our experiments followed standard evaluation methodology and were performed on a desktop system with an AMD Ryzen 9 5900X 12-Core processor.\footnote{\scriptsize{\url{https://www.amd.com/en/products/cpu/amd-ryzen-9-5900x}}}
Our test system ran Ubuntu 22.04.3 LTS, OpenJDK 11.0.20.1, and Python 3.10.13.

\section{Results}

\begin{table}[t]
\centering
\scalebox{0.87}{
\setlength{\tabcolsep}{3pt}
\begin{tabular}{lcccccccc}
\toprule
& \multicolumn{2}{c}{{\bf dev}} & \multicolumn{3}{c}{{\bf DL19}} & \multicolumn{3}{c}{{\bf DL20}}\\
& RR@10 & R@1k & AP & nDCG@10 & R@1k & AP & nDCG@10 & R@1k\\
\cmidrule(lr){2-3} \cmidrule(lr){4-6} \cmidrule(lr){7-9}
BM25 & 0.184 & 0.853 & 0.301 & 0.506 & 0.750 & 0.286 & 0.480 & 0.786 \\
cosDPR-distil & 0.389 & 0.975 & 0.466 & 0.725 & 0.822 & 	0.487 & 0.703 & 0.852 \\
SPLADE++ ED & 0.383 & 0.983 & 0.505 & 0.731 & 0.873 & 0.500 &	0.720 & 0.900 \\
\bottomrule
\end{tabular}
}
\vspace{0.2cm}
\caption{Effectiveness of the learned dense model (cosDPR-distil) and the learned sparse model (SPLADE++ ED) on the MS MARCO passage corpus.}
\label{effectiveness-results}
\end{table}

Effectiveness results are shown in Table~\ref{effectiveness-results}, comparing cosDPR-distil and SPLADE++ ED to baseline BM25 in terms of standard metrics across three sets of queries on the MS MARCO passage corpus.
At a high level, these scores are competitive in terms of retrieval quality to models that are available today, although they remain below the state of the art~\citep{Ma_etal_arXiv2023_RankLLaMA}.
One easy way to increase effectiveness is to perform ensembling of the dense and sparse models (in this case, average of normalized scores), which improves RR@10 on the dev queries to 0.408.
This fusion run is significantly better than both models individually and closer to the state of the art.
However, our main goal here is to provide a point of reference and also to show that dense and sparse retrieval models are roughly on par in terms of effectiveness.

The HNSW index for cosDPR-distil was constructed using the indexer implemented in Anserini, with the parameter \mytt{M} set to 16 and \mytt{efC} set to 1000, using 16 threads, but optimized down to a single index segment (which is a very time-consuming operation).
The cosDPR-distil index is around 27G.
To better understand the impact of different parameter settings, we refer the reader to~\citet{Ma_etal_CIKM2023}.\footnote{Although note that those experiments were conducted with Lucene 9.4.2; the version we used, Lucene 9.8.0, contains additional improvements.}
Due to the non-determinism associated with HNSW indexing, we report scores to three digits on our specific index instance, which are slightly different from the ``reference values'' presented in the Anserini reproducibility documentation.
Effectiveness may vary from run to run, with further dependence on index construction settings (\mytt{M}, \mytt{efC}, and the number of indexing threads).

In contrast, SPLADE++ ED provides deterministic results using an inverted index of only 2.3G, which is about the same size as the inverted index used for BM25, which is 2.6G.
Here, our experiments used an index configuration that only stores impact scores and is also optimized down into a single index segment.
Overall, the sparse index has a smaller memory footprint and can be built faster than the HNSW index.
However, as we will see below, search is slower with inverted indexes than with HNSW indexes.

\begin{table}[t]
\centering
\scalebox{0.87}{
\setlength{\tabcolsep}{3pt}
\begin{tabular}{lrrrr}
\toprule
& \multicolumn{2}{c}{1 thread} & \multicolumn{2}{c}{12 threads} \\
& \multicolumn{2}{c}{throughput (qps)} & \multicolumn{2}{c}{throughput (qps)} \\
& pre-encoded & ONNX & pre-encoded & ONNX \\
\cmidrule(lr){2-3} \cmidrule(lr){4-5}
BM25 & 29.8 $\pm$ 1.2 & - & 278.9 $\pm$ 9.2 & - \\
cosDPR-distil & 27.5 $\pm$ 0.9 & 25.5 $\pm$ 0.3  & 158.0 $\pm$ 0.3 & 72.1 $\pm$ 0.5 \\
SPLADE++ ED & 5.1 $\pm$ 0.2 & 5.0 $\pm$ 0.3 & 48.8 $\pm$ 2.0 & 35.3 $\pm$ 0.2 \\
\bottomrule
\end{tabular}
}
\vspace{0.25cm}
\caption{Performance of Lucene on the MS MARCO dev queries, comparing single-thread query throughput using pre-encoded queries vs.\ ONNX and query throughput on 12 threads using pre-encoded queries vs.\ ONNX.}
\label{performance-results}
\end{table}

Having established the effectiveness of our representative learned dense and sparse retrieval models, we turn our attention to efficiency.
Results are shown in Table~\ref{performance-results}, where we examine two conditions:\
In the first set of experiments, we performed retrieval on a single thread with both models to accurately measure query latency with pre-encoded queries (i.e., cached query representations) and with ``on-the-fly'' query encoding using the ONNX Runtime.
This comparison quantifies the overhead of neural inference for query encoding.
In the second set of experiments, we performed retrieval using 12 threads, fully saturating the CPU of our test machine, to measure throughput.
Once again, we compared pre-encoded queries with on-the-fly query encoding using the ONNX Runtime.

For both sets of experiments, we used the 6980 dev set queries from MS MARCO, retrieving 1000 hits; there are too few TREC queries to obtain reliable measurements.
We quantified performance in terms of throughput, measuring queries per second.
For all conditions, we conducted three warm-up runs to mitigate noise during benchmarking. 
We computed the mean over three subsequent runs, reporting 95\% confidence intervals.
The throughput of BM25 is provided as a point of reference.

These results show that on a single thread, the overhead of query encoding with the ONNX Runtime is relatively modest:\ we see this by comparing ``pre-encoded'' vs.\ ``ONNX'' under the ``1 thread'' condition for both models.
With a single thread, retrieval with HNSW indexes is just slightly slower than retrieval with inverted indexes computing BM25 rankings.
However, retrieval with SPLADE++ ED is much slower than with cosDPR-distil.
We note (again) here that the HNSW indexes have been optimized down to a single segment, which may be an unrealistic assumption for real-world document collections that are not static.

As expected, none of the models exhibit linear scaling (i.e., 12$\times$ more threads does not yield a 12$\times$ increase in throughput), indicating contention in accessing the index structures (under the pre-encoded condition).
The performance gap between BM25 and cosDPR-distil (pre-encoded queries) widens as we scale from 1 to 12 threads, but the gap between cosDPR-distil (pre-encoded) and SPLADE++ ED (pre-encoded) narrows.
These results suggest that, overall, inverted indexes scale better than HNSW indexes under multiple threads.
Interestingly, the gap between ``pre-encoded'' and ``ONNX'' is much larger for 12 threads than for 1 thread, indicating that contention is more severe for ONNX query inference than for accessing index structures.

\section{Discussion}

This paper operationalizes simplicity of design in terms of the minimal number of changes to a baseline architecture that is necessary to gain a certain set of capabilities---in this case, end-to-end retrieval with learned dense and sparse representations.
As previously articulated in~\citet{Lin_etal_arXiv2023_OpenAI}, we begin with the simple observation that search is a brownfield application, which Wikipedia defines as ``a term commonly used in the information technology industry to describe problem spaces needing the development and deployment of new software systems in the immediate presence of existing (legacy) software applications/systems.''
Additionally, ``this implies that any new software architecture must take into account and coexist with live software already in situ.''

That is, most enterprises today already have substantial existing investments in search infrastructure, specifically in Lucene-powered platforms such as Elasticsearch, OpenSearch, and Solr.
The Lucene ecosystem is stable, mature, well supported, and has a massive install base.
In this context, we have shown with Anserini that Lucene coupled with the ONNX Runtime provides all the components necessary to support end-to-end retrieval with the latest dense and sparse models on the CPU, which makes ``bets'' on less mature alternatives (such as dedicated vector stores) less compelling.
Continued investments to grow capabilities in existing Lucene-based stacks are the most likely adoption path we see moving forward.

As advocated in~\citet{devins2022aligning}, alignment between research in neural retrieval and the practice of building search applications brings benefits to both industry and academia (and the research community in general).
For researchers, more rapid technology transfer is an attractive upside, and from this perspective, toolkits such as Anserini and Pyserini help accelerate progress.
Based on our experience, the most important desideratum for researchers is rapid experimental cycles, and thus a pure Java solution (or a Java solution with a thin Python wrapper) can be compelling.
One piece of feedback we have received about the current implementation of Pyserini is its complex dependency chain, which makes installation and configuration a non-trivial barrier for many.\footnote{Python ``dependency hell'' is no doubt a common experience.}
With the capabilities presented here, we see the potential of simplifying first-stage retrieval infrastructure, thereby enabling faster iteration for researchers.

\section{Conclusion}

The dominant narrative today promotes dense retrieval models as a revolutionary advance in search technology that requires entirely new data management infrastructure built around vector stores.
The hype suggests that these models render lexical-based techniques obsolete, and by extension, the venerable inverted index.

Our work pushes back against this narrative.
While transformers have indeed led to significant advances in search effectiveness, the improvements are better characterized as evolutionary.
Furthermore, transformer-based retrieval models are not limited to generating dense representations.
Experimental evidence by many researchers and reproduced here has shown that an entirely different class of learned {\it sparse} models are on par with dense retrieval models in terms of effectiveness.

The truth is that we need {\it both} dense and sparse representations, and that their fusion yields the best results.
However, implementation differences remain, which present practical obstacles to building real-world search systems.
Our position, previously argued in~\citet{Lin_etal_arXiv2023_OpenAI}, is based on the observation that search is a brownfield application, and as a result, many organizations have already made substantial investments that they are unlikely to abandon.
In this context, we believe that the most compelling path forward is the design that requires the fewest changes to existing infrastructure---for most, already built on Lucene---that enables the largest gain in capabilities.
In that sense, Lucene is all you need.

\section*{Acknowledgements}

This research was supported in part by the Natural Sciences and Engineering Research Council (NSERC) of Canada.

\bibliography{onnx}
\bibliographystyle{abbrvnat}

\end{document}